# Interplay of orbital effects and nanoscale strain in topological crystalline insulators

Daniel Walkup[1,2], Badih Assaf[3], Kane L Scipioni[1,4], R. Sankar[5], Fangcheng Chou[5], Guoqing Chang[6], Hsin Lin[7], Ilija Zeljkovic[1,¶] and Vidya Madhavan[4,¥]

[1]Department of Physics, Boston College, Chestnut Hill, Massachusetts 02467, USA

[2] National Institute of Standards and Technology, Gaithersburg, Maryland 20899, USA

[3] Département de Physique, Ecole Normale Supérieure, CNRS, PSL Research University, 75005 Paris, France

[4]Department of Physics and Frederick Seitz Materials Research Laboratory, University of Illinois Urbana-Champaign, Urbana, Illinois 61801, USA

[5]Center for Condensed Matter Sciences, National Taiwan University, Taipei 10617, Taiwan

[6]Centre for Advanced 2D Materials and Graphene Research Centre, National University of Singapore, Singapore 117546

[7]Department of Physics, National University of Singapore, Singapore 117542

[¶] co-corresponding author (ilija.zeljkovic@bc.edu)

[¥] co-corresponding author (vm1@illinois.edu)

## Abstract

**Orbital degrees of freedom can have pronounced effects on the fundamental properties of electrons in solids. In addition to influencing bandwidths, gaps, correlation strength and dispersion, orbital effects have also been implicated in generating novel electronic and structural phases, such as Jahn-Teller effect and colossal magnetoresistance. In this work, we show for the first time how the orbital nature of bands can result in non-trivial effects of strain on the band structure. We use scanning tunneling microscopy and quasiparticle interference imaging to study the effects of strain on the electronic structure of a heteroepitaxial thin film of a topological crystalline insulator, SnTe. We find a surprising effect where strain applied in one direction affects the band structure in the perpendicular direction. Our theoretical calculations indicate that this effect directly arises from the orbital nature of the conduction and valance bands. Our results imply that a microscopic model capturing strain effects on the band structure must include a consideration of the orbital nature of the bands.**

Topological crystalline insulators (TCIs) are a recently discovered [1,2] subclass of 3D topological materials which harbor massless Dirac surface states (SS) tunable by temperature [3,4] and alloying composition change [5]. In contrast to $Z_2$ topological insulators [6–8] in which the Dirac crossing is protected by time-reversal symmetry, Dirac point in TCIs is protected by a discrete set of crystalline symmetries [1]. This unique coupling between the crystal structure and the Dirac SS provides a route towards controlling the SS dispersion by using different types of structural deformations. Theory predicted [9] and experiments confirmed [10] that a lattice distortion that breaks the mirror symmetry protecting the Dirac point in TCIs enables otherwise massless Dirac SS fermions to acquire mass. However, from both the fundamental and the applications perspectives, one of the key goals remains uncovering new pathways for the manipulation of topological SS via structural deformations without breaking any crystalline symmetry protecting the Dirac nodes.

Theoretically, strain in TCIs is predicted to give rise to a variety of exotic phenomena such as pseudomagnetic fields, quantum phase transition from the trivial to the topological state, momentum-space evolution of the Dirac nodes and unconventional superconductivity [11,12]. The challenge in achieving many of these remains the difficulty of controllably applying strain at the surface of TCIs, characterizing the type of strain induced, quantifying its magnitude and simultaneously measuring the electronic band structure. Theory predicts that different types of structural distortions will lead to a distinct behavior of the SS band structure [11]. For example, on the (001) face of TCI (Pb,Sn)(Se,Te), which hosts a Dirac cone near each X point, strain applied equally along both in-plane lattice directions is expected to lead to the symmetric shift of all four Dirac nodes within the first Brillouin zone. Strain applied exclusively along a single lattice direction on the other hand is expected to selectively tune the momentum space position two Dirac nodes, while shear strain could, in addition to independent tunability of the Dirac nodes, also break crystalline symmetry and create massive Dirac fermions. Our recent experiments have probed the effects of biaxial strain on the Dirac nodes [13], however, many questions remain on the effects of uniaxial and sheer strains on the band structure of this Dirac system.

It has been known for decades that many (001)-oriented heterostructures of IV-VI semiconductors exhibit grid-like quasi-periodic arrays of misfit dislocations, associated with strong strain patterns near the interface.[13–15] These sub-surface dislocations manifest themselves as linear "dips" or troughs in the STM topographs directly above the line of the dislocation[14,16] (Fig. 1). In a previous study, we showed that the ~3% lattice mismatch between TCI SnTe and non-topological insulator PbSe (001) substrates generates a spatially inhomogeneous strain field and we were able to characterize the effects of biaxial strain in this system. In this work we analyze high-resolution scanning tunneling microscopy (STM) topographs of SnTe/PbSe(001) heteroepitaxial structure to discover that misfit dislocations induce regions of both biaxial, uniaxial and sheer strain. Since the shear strain component is small, we focus on the effects of biaxial and uniaxial strains and use simultaneous mapping of the electronic structure via quasiparticle interference (QPI) imaging over the identical region of the sample to determine the effects of these types of strain. In contrast to our previous studies of biaxial strain, uniaxial strain allows us to distinguish the strain along the two axes, and determine the effect of each on the Dirac nodes. Surprisingly, our experiments show that strain applied in one direction has the most pronounced effect on the Dirac cones lying along *the perpendicular* direction. Our theoretical calculations indicate that this counter-intuitive effect can be directly attributed to changes in the overlap of the various *p*-orbitals at adjacent atomic sites, as the distances are modulated by strain.

We first quantify different types of strain at the surface of our films (Fig. 1). Starting with an atomically resolved STM topograph $T(r)$ (Fig. 2(a)), we apply the Lawler-Fujita drift correction algorithm with small drift-correction length scale of 2-3 lattice constants [17]. This algorithm applies a transformation $r \rightarrow r + u(r)$ to the topograph, such that the transformed topograph $T'(r)$ contains a perfectly periodic atomic lattice. The normalized displacement field $u(r)$ can be viewed as the displacement vector in elasticity theory [18], whose derivatives give the strain tensor. Individual strain tensor components extracted from our data (Figs. 2(c-f)) enable us to characterize the type and local magnitude of strain as follows.

Using strain tensor components $u_{xx}(r)$, $u_{xy}(r)$, $u_{yx}(r)$ and $u_{yy}(r)$, we can extract the spatially varying magnitude of: (1) biaxial strain as $C \equiv (u_{xx}(r) + u_{yy}(r))/2$ (Fig. 2(g)), (2) uniaxial strain as $U \equiv (u_{xx}(r)-u_{yy}(r))/2$ (Fig. 2(h)), (3) shear strain as $S \equiv (u_{xy}(r)+u_{yx}(r))/2$ (Supplementary Information I), and (4) local rotation of the lattice $R \equiv (u_{xy}(r)-u_{yx}(r))/2$ (Supplementary Information I). The dominant strain types measured in our thin film are biaxial (Fig. 2(g)) and uniaxial (Fig. 2(h)) strain, while the shear strain and the local lattice rotation magnitudes are significantly weaker (Supplementary Information I). The diagonal elements of the strain tensor (Figs. 2 (c,f)) each show a clear one-dimensional pattern, with blue lines of compression co-located with the troughs in the topograph (Fig. 2 (a)). The orientation of the pattern is consistent with its origin in the network of misfit dislocations at the interface. The derived quantities, C and U, each show strong variation determined by their basic constituents. The compression (Fig. 2(g)) is greatest where two troughs intersect, and smallest (negative) in the midpoint between intersections, producing a distinctive cell pattern which resembles the topograph in Fig. 2(a). The uniaxial stretch (Fig. 2(i)) is greatest where maxima of $u_{xx}(r)$ coincide with minima of $u_{yy}(r)$.

Next, we proceed to measure the Dirac surface state band structure, by using the quasiparticle interference (QPI) imaging method [19]. QPI imaging relies on the elastic scattering of quasiparticles on the surface of a material, which produce standing waves in the density of states. These standing waves appear as oscillations or ripples in the measured local density of state, $G(r,V) \equiv dI/dV(r,V)$, with wavevector $\mathbf{q} = \mathbf{k_i} - \mathbf{k_f}$, where $\mathbf{k_{i,f}}$ are the initial and final momenta of the scattered quasiparticles. These **q**-vectors can be directly extracted from the Fourier transform of G(r,V), and reveal the momentum-space position of the underlying surface states.

The surface states of SnTe(001) consist of a pair of Dirac cones near each X and Y points. Each pair undergoes a Lifshitz transition so that at the energies shown here (~200-250 meV below the Dirac point), the constant-energy-contours resemble the ellipses around each X and Y points (inset of Fig. 3(a)).[2,20] In this work, we will focus on the vectors labeled $Q_{1x}$ and $Q_{1y}$ in Fig. 3(a). Each one corresponds to a single "valley" of Dirac fermions, and represents scattering across the Γ point between two ellipses on opposite sides.

Before discussing the effects of biaxial and uniaxial strain, which affect both lattice directions, let us first focus on a simple case in which the lattice constant is modulated in only one direction, for example along x-axis (Fig. 1c and Fig. 2c). To determine the change in the Dirac surface state dispersion in response to this type of strain, we separate G(r,V) into a series of stripes based on $u_{xx}$ map acquired over the same area of the sample (Fig. 2c). This is done by masking the G(r,V) into separate areas $G_m^i(r,V)$, with each $G_m^i(r,V)$ consisting of a fraction of the total area of the image. These areas are chosen such that they can be assigned a single, average value of strain measured within the mask. In this scenario, the mask can be most easily visualized as a series of one-dimensional strips parallel to y (Supplemental Information II). To determine the average Dirac surface state structure within each mask, we apply a two-dimensional Fourier-transform (FT) to each $G_m^i(r,V)$, and examine the QPI wavevectors (Figs. 3b), similarly to band mapping of TCIs done previously [13,21]. Qualitatively, it is sufficient to compare FTs of $G_m^i(r,V)$ for the two masks capturing the extreme values of each type of strain – the mask encompassing the area of the largest compressive strain and the mask consisting of the area with the largest tensile strain (Fig. 3b). Interestingly, we can immediately observe that the strain applied along x causes most prominent change in the orthogonal direction, along y, by shifting the position of the $Q_{1y}$. Conversely, the strain applied along y causes most prominent change in the orthogonal direction, by shifting the position of $Q_{1x}$.

Using the equivalent masking procedure described in the previous paragraph, we proceed to investigate and discuss the effects of biaxial and uniaxial strain. Compressive biaxial strain causes both $Q_{1x}$ and $Q_{1y}$ to shift towards the center of the FT by approximately the same amount (Fig. 3d), as expected based on theory [11]. This reflects the stretching of the ellipse-pairs at X and Y respectively, which is rooted in the momentum space shift of the Dirac cones towards Γ. Uniaxial strain on the other hand induces the shift in $Q_{1x}$ and $Q_{1y}$ in the opposite directions (Fig. 3e). To extract a quantitative shift of the Dirac nodes in response to each type of strain, we plot the evolution of $\Delta Q_{1x}$ and $\Delta Q_{1y}$ along both lattice directions as a function of strain (Fig. 4). This analysis confirms our surprising discovery that the strain applied along one lattice direction results in the most prominent shift of the Dirac nodes along the perpendicular direction (Figs. 4a,b). This leads us to the question: what is the origin of this counter-intuitive response of the Dirac cones in response to uniaxial strain? To answer this question, we carried out a series of calculations as described below.

In general, momentum-space position of the Dirac nodes around each X point in TCIs is directly related to the bulk band gap at X – the larger the bulk band gap, the further the Dirac node is from X. Therefore, to determine what drives the evolution of Dirac nodes, we need to consider the effects influencing the magnitude of the bulk gap. Theoretically, we can consider a two-center approximation in tight-binding formalism, with $t_1$ and $t_2$ hopping terms between adjacent Te atomic sites shown in Fig. 5a. Since conduction bands at X and Y points are mainly composed from Te atoms and valance bands of X and Y points come from Sn, $t_1$ and $t_2$ hopping terms will only shift the energy level of the conduction bands. The conduction bands at X are mainly composed from $p_x$ orbitals, while those at Y mainly consist from $p_y$ orbitals (Fig. 5). To compare the band gap change at X and Y, we only need to know the change of $t_1$ and $t_2$. When y-axis is squeezed, one can show that the change of $t_2$ is greater than that of $t_1$ by replacing y by y-dy and calculating the leading term of dx (Supplementary Information III). Qualitatively, this can be seen by observing the overlap of orbitals as the lattice is strained (Fig. 5c). In this scenario, the "overlap" between $p_y$ and $p_z$ orbitals will not significantly change, but the overlap between $p_x$ and $p_z$ is clearly magnified. Thus, strain along y-axis will not change the bulk band gap at Y, but will cause the change of the bulk band gap at X. This in turn implies that the Dirac node shift of the Dirac cones near X will be significantly stronger than the shift of the Dirac cones near Y (Fig. 5d), which is exactly confirmed in our experiments. Our results have important implications for creating microscopic models for strain effects on band structure and imply that any microscopic strain model must include a consideration of the orbital nature of the bands.

## Acknowledgements

V.M gratefully acknowledges support from the U.S. Department of Energy (DOE), Scanned Probe Division under award DE-SC0014335 for STM studies and the support of D W. and K.L. S., and the National Science Foundation award NSF-ECCS-1232105 for thin film growth and for the support B.A.A. B.A.A. is also partially supported by ANR LabEx grant ENS-ICFP (ANR-10-LABX-0010/ANR-10-IDEX-0001-02 PSL).

**Methods**

Thin film used in our experiment was synthesized using e-beam evaporation method in an ultra-high vacuum chamber directly attached to the STM. SnTe thin films were deposited at 300 °C at the growth rate of ~ 2 monolayers (ML) per minute for a total ~40 ML thickness as determined by the thickness of SnTe thin films grown on Si(001) under the same conditions. PbSe single crystal used as a substrate was grown by the self-selecting vapor growth method and cleaved at room-temperature in ultra-high vacuum along the (001) direction to expose a pristine surface free of contaminants. Within minutes after the completion of the growth process, the films were inserted into the STM head where they were held at 4K during the duration of the experiments presented in this work. All STM data was acquired at ~4.5 Kelvin using a Unisoku USM1300 STM. dI/dV maps were acquired using 10 meV lock-in excitation amplitude and 1488 Hz frequency.

# Figures

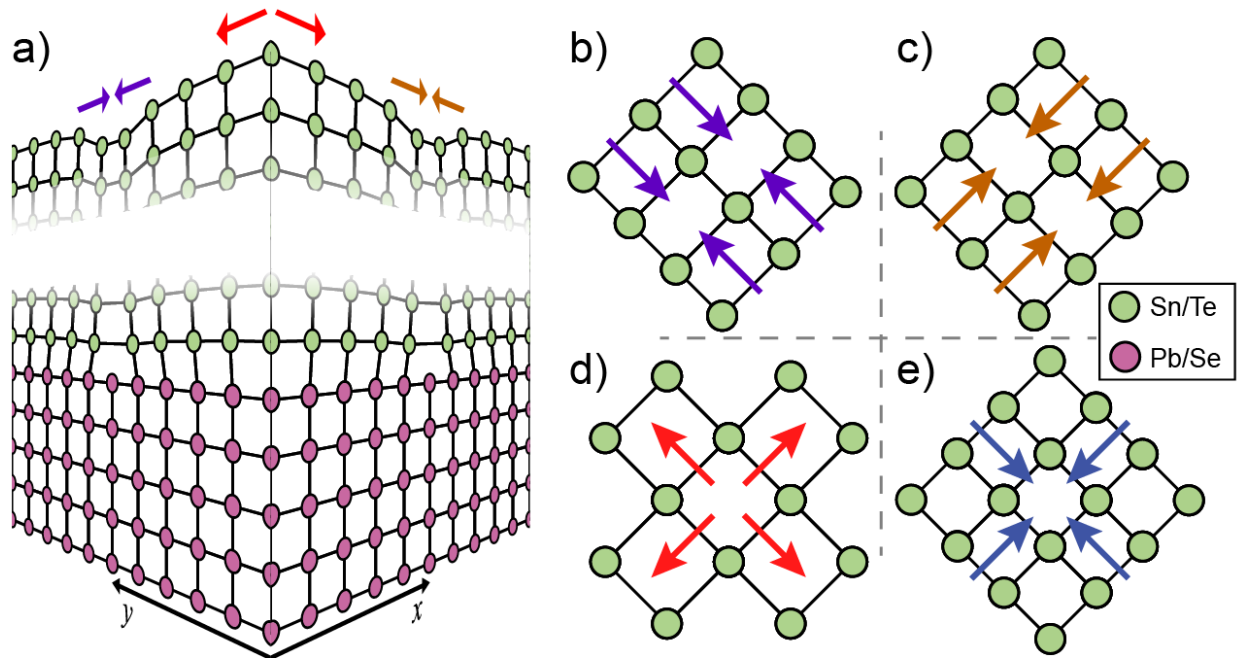

**Figure 1. Different types of strain. (a)** Schematic of the misfit dislocation network appearing at the interface between the PbSe (001) substrate (purple) and the SnTe film (green). Lattice distortions in the topmost SnTe film in the presence of **(b)** uniaxial compression along y-axis, **(c)** uniaxial compression along x-axis, **(d)** biaxial tensile strain and **(e)** biaxial compressive strain. The color of arrows in (b-e) indicates the position on the sample in (a).

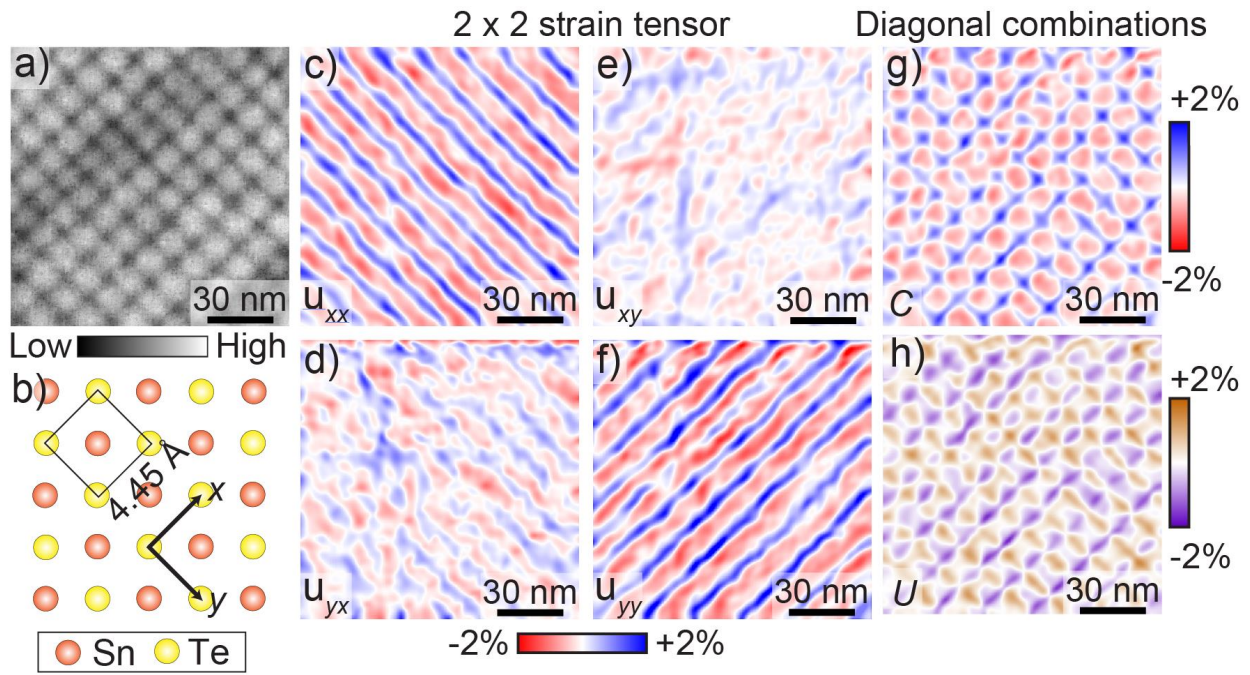

**Figure 2: Spatial distribution of different types of strain. (a)** STM topograph of ~130 nm square region of the sample ($V_{set}$ = -50 mV, $I_{set}$ = 200 pA) **(b)** Schematic of the (001) surface of SnTe. Arrows in (b) denote the x- and y-axes. **(c)-(f)** The components of the 2 x 2 strain tensor $\nabla \mathbf{u}(\mathbf{r})$ extracted from topograph in **(a)**. $u_{ij}$ denotes $\frac{\partial u_i}{\partial j}$. **(g)** The biaxial strain map C. **(i)** The uniaxial strain map U.

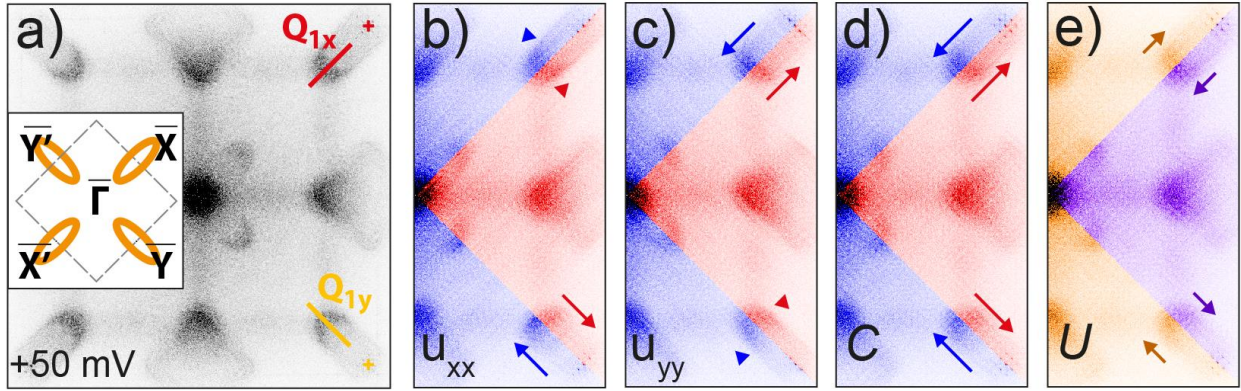

**Figure 3. Strain-filtered Fourier transforms of dI/dV. (a)** The Fourier transform of dI/dV acquired in the 130nm area shown in Fig. 2(a); the indicated features **Q**$_{1x}$ and **Q**$_{1y}$ represent scattering across the center of the Brillouin zone between the inner portions of the pockets (inset) at $\bar{X}$, $\bar{X}'$, and $\bar{Y}$, $\bar{Y}'$ respectively. **(b)-(e)** The Fourier transforms of masked dI/dV, with masks chosen to capture the maxima (blue in **(b)-(d)**, orange in **(e)**), and minima (red in **(b)-(d)** and purple in **(e)**) of $u_{xx}$, $u_{yy}$, $C$, and $U$ respectively; the arrows are guides to the eye. In **(b)**, **(c)**, **(d)** (**(e)**) the blue (orange) and red (purple) subsets correspond to masks coincident with the blue (orange) and red (purple) areas of Fig. 2 (c), (f), (g), and ((i)) respectively. For details of the masking procedure, see Supplementary Information I.

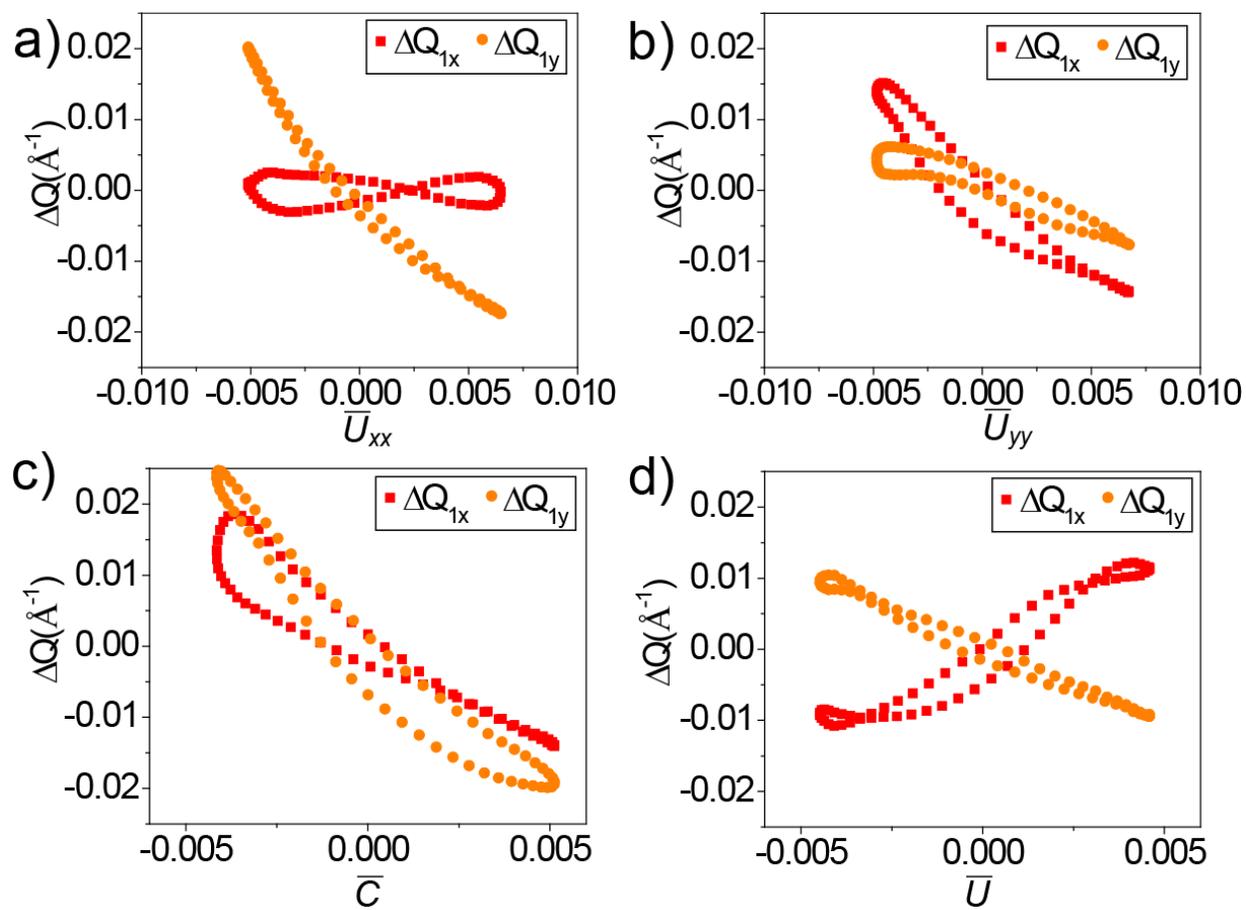

**Figure 4. The shift of the QPI peaks as a function of average strain.** ΔQ$_{1x}$ and ΔQ$_{1y}$ shifts plotted against the average: **(a)** u$_{xx}$, **(b)** u$_{yy}$, **(c)** C and **(d)** U.

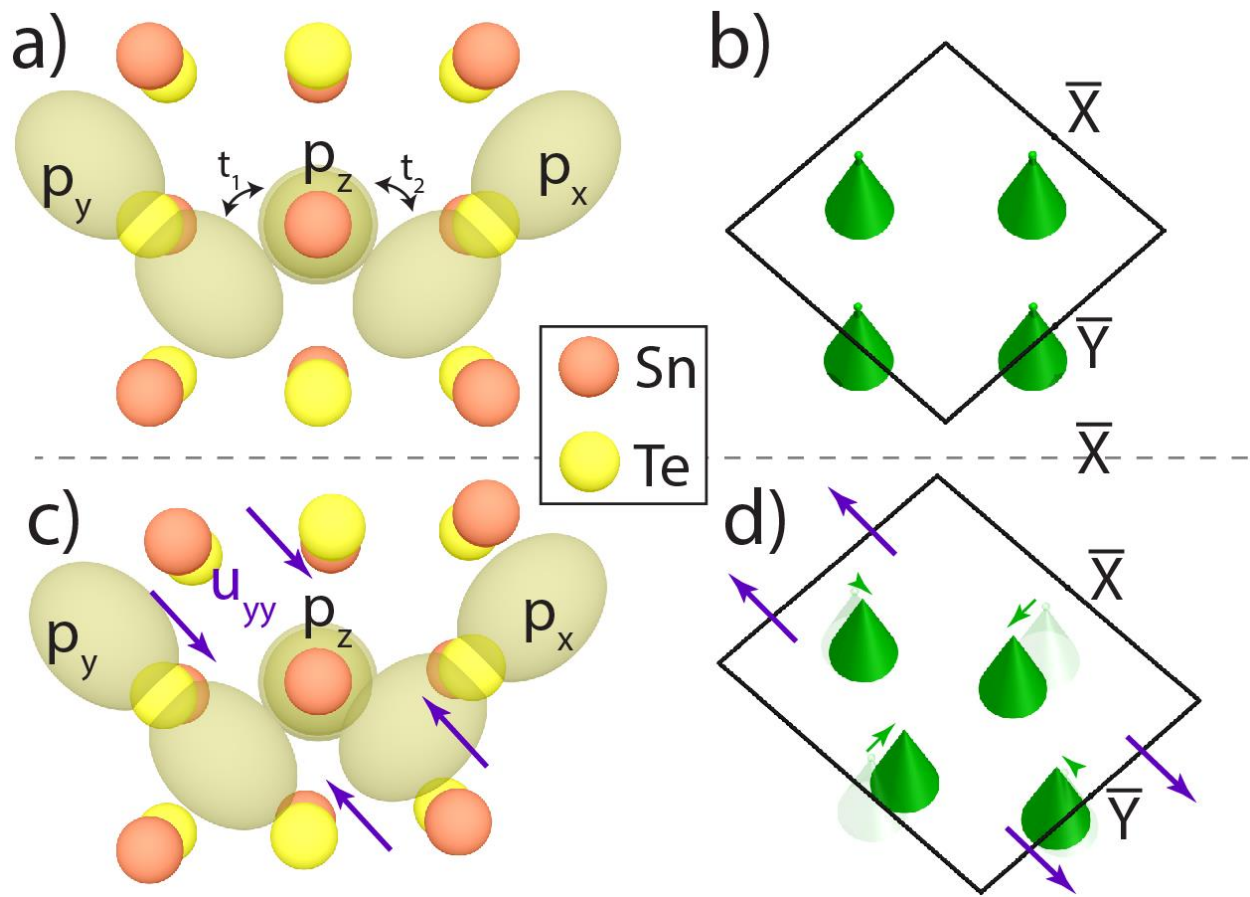

**Figure 5. The effects of orbital overlap on the electronic band structure in TCIs.** Schematic of the relevant orbitals: **(a)** in the absence of any distortion, and **(c)** under strain along y-axis. **(b,d)** The positions of the four Dirac cones within the first Brillouin zone related to the atomic structure in (a) and (c), respectively. As seen in panel (c), strain in y-direction increases the "overlap" of $p_x$ and $p_z$ orbitals, while the overlap of $p_y$ and $p_z$ does not significantly change.